\documentstyle[preprint,aps,prl,epsf]{revtex}
\tightenlines
\def\zbbar{$Z^0\rightarrow b{\overline b}$}

\def\bbr{{\overline b}}

\def\htb{{\hat T}}

\begin{document}


 \begin{flushright}
 SLAC--PUB--9188   \\
 \vspace*{-3mm}
 SCIPP--02/18      \\
 \vspace*{-3mm}
 August 2002
 \end{flushright}

\begin{center}
\begin{large}
{\bf Improved Direct Measurement of the Parity-Violation
 Parameter \\}
\vspace*{-4mm}
{\bf $A_{b}$ Using a Mass Tag and Momentum-Weighted Track Charge   }

\vspace*{8mm}
{\bf The SLD Collaboration$^{**}$}
\end{large}

\vspace*{2mm}

Stanford Linear Accelerator Center    \\
\vspace*{-2mm}
Stanford University, Stanford, CA 94309

\vspace*{12mm}
{\bf Abstract}
\end{center}


We present an improved direct measurement of the parity-violation parameter
$A_b$ in the $Z$ boson--b quark coupling
using a self-calibrating track-charge technique applied to a sample enriched in
$Z \rightarrow b {\bar b}$ events via the topological reconstruction of the B hadron
mass. Manipulation of the SLC electron-beam polarization permits the
measurement of $A_b$ to be made independently of other $Z$-pole coupling parameters.
From the 1996-98 sample of 
400,000 hadronic $Z$ decays,
produced with an average
beam polarization of 73.4\%, we find $A_b = 0.906 \pm 0.022({\rm stat.}) \pm
0.023({\rm syst.})$. 



\vspace*{28mm}
\begin{center}
{\it Submitted to Physical Review Letters}
\end{center}

$^*$ Work supported in part by Department of Energy contract 
DE-AC03-76SF00515

\vfill
\eject

\narrowtext

Measurements of $b$ quark production asymmetries at the $Z^0$ pole 
determine the extent of parity violation in the $Zb\bar{b}$ coupling.
At Born level, the differential cross section for 
the process $e^+e^-\rightarrow Z^0 \rightarrow b\bar{b}$
can be expressed 
as a function of the polar angle  
$\theta$ of the $b$ quark
relative to the electron beam direction, 
\begin{equation}
\label{DIFFXSECT}
\sigma^b(\cos\theta) \equiv
d\sigma_b/d\cos\theta
 \propto (1 - A_e P_e)(1 + \cos^2 \theta) +
2 A_b (A_e - P_e) \cos \theta ,\nonumber
\end{equation}
where
$P_e$ is the longitudinal polarization of the electron beam  ($P_e > 0$ for
predominantly right-handed polarized beam).
The parameter $A_f = 2v_f a_f/(v_f^2+a_f^2)$, where
$v_f\ (a_f)$ is the vector (axial vector) coupling of the fermion $f$
to the $Z^0$ boson, with $f = e$ or $b$,
expresses the extent of parity violation in the $Zf\bar{f}$ coupling.

From the conventional forward-backward asymmetries formed with an unpolarized electron beam
($P_e = 0$),
such as that used by the CERN Large Electron-Positron Collider (LEP) experiments,
only the product $A_eA_b$ of parity-violation parameters
can be measured ~\cite{LEPEWWG}.
With a longitudinally polarized electron beam, however, it is possible to measure
$A_b$ independently of $A_e$ by fitting simultaneously to the differential
cross sections of Eq. (1) formed separately for predominantly left-
and right-handed beam. The resulting direct measurement of $A_b$ is
largely independent of
propagator effects that modify the effective weak mixing angle,
and thus is complementary to other electroweak
asymmetry measurements performed at the $Z^0$ pole.

In this Letter, we present a measurement of $A_b$ based on the
use of an inclusive vertex mass tag (improved relative to that of
previous publications due to the use of an upgraded vertex
detector)
to select $Z \rightarrow b {\bar b}$
events, and the net momentum-weighted track charge~\cite{feyf}
to identify the charge of the
underlying quark.
This result, incorporating data collected
during the 1996-98 runs of the Stanford Linear
Collider (SLC), is over twice as precise as that of our
previous publication~\cite{prl98}, which was based on data from 1993-95.

The operation of the SLC with a polarized electron beam has been described
elsewhere~\cite{SLCpol}. During the 1996-98 run, the SLC Large Detector (SLD)~\cite{SLD}
recorded an integrated luminosity of 14.0 pb$^{-1}$, at a mean center-of-mass
energy of 91.24 GeV, and with a luminosity-weighted mean electron-beam polarization
of $|P_e| = 0.734 \pm 0.004$~\cite{alr}.
The 1996-98 run of the SLD detector incorporated the upgraded VXD3
CCD pixel vertex detector~\cite{vxd3}, which featured a greater coverage in
$\cos \theta$, as well as a larger outer radius and substantially less material
per layer, than that of the VXD2 vertex detector~\cite{vxd2}
in place from 1993-95.

The SLD measures charged particle tracks
with the Central Drift Chamber (CDC), which is immersed in a
uniform axial magnetic field of 0.6T.
The VXD3 vertex detector
provides an accurate measure
of particle trajectories close to the beam axis.
For the 1996-98 data, the combined $r\phi$ ($rz$)
impact parameter resolution of
the CDC and VXD3 is
7.7 (9.6) $\mu$m at high momentum, and 34 (34) $\mu$m at
$p_{\perp}\sqrt{\sin \theta}$ = 1 GeV/c, where
$p_{\perp}$ is the momentum transverse to the beam direction, and
$r$ ($z$) is the coordinate perpendicular (parallel)
to the beam axis.
The combined momentum resolution in the plane
perpendicular to the beam axis is
$ \delta p_{\bot} / p_{\bot} =
\sqrt{(.01)^2+(.0026~p_{\bot}/GeV/c)^2 \,}$.
The thrust axis is reconstructed using the Liquid Argon
Calorimeter, which covers the angular range $|\cos \theta| < 0.98$.

The details of the analysis procedure are similar to
those of the 1993-95 sample analysis.
Events are classified as hadronic  $Z^0$ decays if
they: (1) contain at least seven well-measured tracks
(as described in Ref.~\cite{SLD});
(2) exhibit a visible charged energy of at least 20 GeV;
and (3) have a thrust axis polar angle satisfying
$|\cos\theta_{thrust}| < 0.7$. The resulting hadronic sample 
from the 1996-98 data consists of 245,048 events
with a non-hadronic background estimated to be $< 0.1 \%$.

We select against multi-jet events in order to reduce the
dependence of the measured value of $A_b$ on the effects of
gluon radiation and inter-hemisphere correlation.
Events are discarded if they are found to have four or more
jets by
the  JADE jet-finding
algorithm with $y_{cut} = 0.02$ \cite{JADE}, 
using reconstructed charged tracks as input.
In addition, any event found to have three or
more jets with $y_{cut} = 0.1$ is discarded.

To increase the \zbbar\ content of the sample, a tagging procedure
based on the invariant mass of
3-dimensional topologically
reconstructed secondary decay vertices is applied \cite{ZVTOP}.
The mass of the reconstructed vertex is
corrected for missing transverse
momentum relative to the reconstructed $B$ hadron flight
direction in order to partially account
for neutral particles. The requirement that the event contain at
least one
secondary vertex with mass greater than
2 GeV/c$^2$ results in a sample of 36,936 candidate \zbbar\
decays. 
The purity (97\%) and efficiency (77\%) of this sample
are calculated from the data by comparing the rates for
finding a high mass vertex in either a single or both hemispheres,
where the two hemispheres are defined relative to the plane perpendicular
to the thrust axis.
This procedure assumes {\it a-priori} knowledge of the small
$udsc$ tagging efficiency, as well as the size of inter-hemisphere
correlations, both of which are taken from Monte Carlo (MC) simulation.
This procedure also assumes knowledge of the $Z \rightarrow
c \bar{c}$ and $Z \rightarrow b \bar{b}$ branching fractions,
which are assigned their Standard Model
values of 0.172 and 0.216, respectively. 
 
\begin{figure}[htb]
\bigskip
\epsfxsize=5.3in
\epsfysize=5.0in
\epsfbox{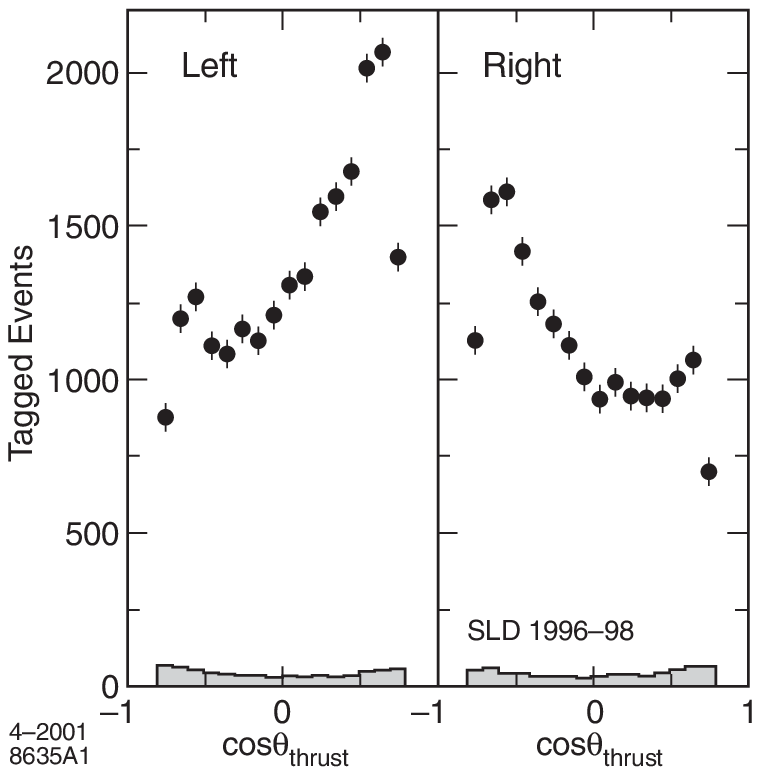}
\caption{Polar angle distributions for track-charge-signed $Z \rightarrow
b \bar{b}$ candidates, separately for left- and right-handed electron beam.
The shaded histogram represents the contribution from non-$b {\bar b}$
background, estimated as described in the text. The analysis employs a cut
of $|\cos\theta| < 0.7$.}
\label{POLARANG}
\end{figure}

We construct a signed thrust axis $\htb$, which provides an estimate
of the direction of the negatively-charged $b$ quark, as follows.
Using all track-charge quality tracks, as defined in Ref.~\cite{JC94},
we form the 
track-direction-signed ($Q$) and unsigned ($Q_+$) momentum-weighted
track-charge sums
\begin{equation}
\label{PWCH}
Q = -\sum_{tracks}q_j \cdot \mbox{sgn}(\vec{p}_j \cdot \htb )
  |(\vec{p}_j \cdot \htb )|^{\kappa},
\end{equation}
\begin{equation}
\label{PWCHS}
Q_+ = \sum_{tracks}q_j|(\vec{p}_j \cdot \htb )|^{\kappa},
\end{equation}
where $q_j$ and $\vec{p}_j$ are the  charge and momentum of track $j$, 
respectively. $\htb$ is chosen as the unit vector parallel to the thrust axis
that renders $Q > 0$.
We use $\kappa = 0.5$ to maximize the analyzing power
of the track charge algorithm for \zbbar\ events, resulting
in a correct-assignment probability of 70\%.
Fig.~\ref{POLARANG} shows the
$T_z=\cos\theta_{thrust}$
distribution of the $b$-enriched sample separately
for left- and right-handed electron beams. Clear forward-backward
asymmetries are observed, with respective signs as expected
from the cross-section formula in Eq. \protect\ref{DIFFXSECT}.

The value of $A_b$ is extracted via a maximum likelihood 
fit to the differential
cross section (see Eq. \ref{DIFFXSECT})
\begin{eqnarray}
\label{LIKELIHOODFUNCT}
 \rho^i (A_b) = (1-A_eP_e^i)(1+(T_z^i)^2)
 + 2(A_e-P_e^i)T_z^i
  [A_bf^i_b(2p_{b}^i-1)&(1-\Delta_{QCD,b}^i) + \nonumber\\
A_cf^i_c(2p_{c}^i-1)(1-\Delta_{QCD,c}^i) +
A_{bckg}(1-f^i_b-f^i_c)(2p_{bckg}^i-1)],&
\end{eqnarray}
where $P_e^i$ is the signed polarization of the electron beam for 
event $i$, $f_{b(c)}^i$ 
the probability
that the event is a $Z^0\rightarrow b{\overline b}(c{\overline c})$
decay (parameterized as a function of
the secondary vertex mass),
and $\Delta_{QCD,b,c}^i$ are final-state
QCD corrections, to be discussed below.
$A_{bckg}$ is the estimated 
asymmetry of residual
$u\overline u$, $d\overline d$, and $s \overline s$ final states.
The parameters $p_{ }$ are estimates of the probability
that the sign of $Q$ accurately reflects the charge of the
respective underlying quark, and are functions of $|Q|$, as well as the 
secondary vertex mass and $|T_z|$.

\begin{figure}[htb]
\epsfxsize 5.0in
\epsfysize 5.0in
\epsfbox{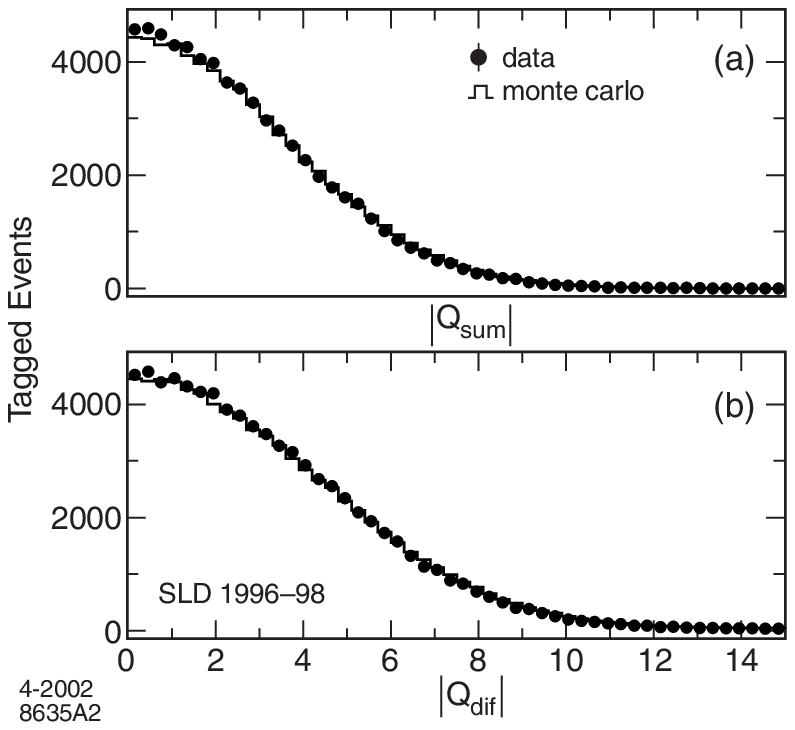}
\caption{Comparison between data (points) and MC (histogram) for the
observables $|Q_{sum}|$ and $|Q_{dif}|$ (see text),
for $Z \rightarrow b \bar{b}$ candidates.}
\label{QDIST}
\end{figure}

As in our previous publication \cite{prl98},
we measure $p_{b}$ directly from the data \cite{THESIS}.
Defining $Q_b$ ($Q_\bbr$) to be the track-direction-unsigned
momentum-weighted track charge
sum for the thrust hemisphere containing
the $b$ ($\bbr$) quark,
the quantities
\begin{equation}
Q_{sum} = Q_b + Q_\bbr\ ,\ \ \ Q_{dif} = Q_b - Q_\bbr\ ,
\end{equation}
may be related to the experimental observables defined in Eqs.~\ref{PWCH}
and \ref{PWCHS} respectively: $|Q_{dif}|=|Q|$ and $Q_{sum} = Q_+$ .
Our MC simulation indicates that the $Q_b$ and
$Q_\bbr$ distributions are approximately Gaussian.
In this limit \cite{THESIS},

\begin{equation}
p_{b}(|Q|) = \frac{1}{1+e^{-\alpha_b|Q|}}\ ,
\end{equation}
with
\begin{equation}
\alpha_b = \frac{2q_{dif}^0}{\sigma_{dif}^2} = 
  \frac{2\sqrt{\langle {\left |Q_{dif}\right |}^2 \rangle -
\sigma_{dif}^2}}{\sigma_{dif}^2}\ ,
\end{equation}
where
$q_{dif}^0$ and $\sigma_{dif}$ are the mean and width, respectively,
of the Gaussian $Q_{dif}$ distribution. 
The parameter $\alpha_b$, whose magnitude depends upon the separation
between the $b$ and $\bbr$ track-sum distributions via the observable
$\langle {\left |Q_{dif}\right |}^2 \rangle$, provides a measure of the
analyzing power of the $b$-quark direction estimator $\htb$.
Figure~\ref{QDIST} compares
the distributions
of the observable combinations $\left | Q_{dif} \right |$ and $Q_+$
between data and MC.

In the absence of a correlation between $Q_b$ and $Q_\bbr$,
$\sigma_{dif} = \sigma_{sum}$, where $\sigma_{sum}$ is the observed
width of the $Q_{+}$ distribution.
Thus $\alpha_b$ can be derived from 
experimental observables.
In the presence of a correlation, 
$\sigma_{dif} = (1+\lambda)\sigma_{sum}$,
where $\lambda$ characterizes the strength of the correlation,
which can be determined from the MC simulation.
For JETSET 7.4 \cite{SJO87} with parton shower evolution, string
fragmentation, and full detector simulation,
$\lambda$ is found to be 0.040.
The effects of light flavor contamination are taken into
account by adjusting the observed widths $\sigma_{sum}^2$
and $\langle \left |Q_{dif} \right |^2 \rangle$,
using the magnitude and width of the light-flavor and $c\bar{c}$
contributions estimated from the MC. This correction increases
the value of $\alpha_b$ by 2\% to $0.2944 \pm 0.0078$,
bringing it into good agreement with
the value of $0.2949 \pm 0.0007$
extracted from the $Z \rightarrow b {\bar b}$
simulation.

Final-state gluon radiation reduces the observed asymmetry
from its Born-level value. 
 This effect is incorporated in our analysis by applying a correction
$\Delta_{QCD}(|\cos\theta|)$
to the likelihood function 
(Eq.~\ref{LIKELIHOODFUNCT}).
Calculation of the quantity $\Delta_{QCD}$ has been performed by several
groups~\cite{QCDall}. 

For an unbiased sample of $b {\bar b}$ events,
correcting for final-state gluon radiation increases the measured
asymmetry by $\sim 3\%$.
However, QCD radiative effects are mitigated
by the use of the thrust axis to estimate the $b$-quark direction, the
\zbbar\ enrichment algorithm, the self-calibration procedure,
and the cut on the number of jets.
A MC simulation of the analysis chain indicates that these effects can
be represented by a $\cos\theta$-independent suppression factor,
$x_{QCD}=0.074$, such that
$\Delta_{QCD} =  x_{QCD} \Delta^{TH}_{QCD}$.

Effects due to gluon splitting to $b {\bar b}$ and $c {\bar c}$
have been estimated by rescaling the JETSET simulation production of
such quark pairs to current world-average gluon splitting measurements~\cite{split},
leading to a correction of +0.3\% on the value of $A_b$.
Additional radiative effects, such as those due to initial-state radiation
and $\gamma/Z$ interference, lead to a further correction of $-0.2\%$
to the measured value of $A_b$.

While, as described above, the overall tagging efficiency is derived
from data,
the dependence of the $b$-tagging efficiency
upon the secondary vertex mass must be estimated from
the MC simulation, as must be the
charm correct-signing probability $p_{c}$.  The value of
$A_c$ is set to its Standard Model value of 0.67, with an uncertainty
commensurate with that of~\cite{CERNEWWG}.
The value
of $A_{bckg}$ is set to zero, with an uncertainty corresponding to
the full physical range $|A_{bckg}| < 1$.
The resulting value of $A_b$ extracted from the fit
is $A_b=0.907\pm0.022\,\,(stat)$.
This result is found to be insensitive to the value of the 
$b$-tag mass cut, and the value of weighting exponent $\kappa$ used in the
definition (\ref{PWCH}) and (\ref{PWCHS}) of the momentum-weighted track charge sum.

We have investigated a number of systematic effects which can change the
measured value of $A_b$; these are summarized in Table~\ref{SYSERR}.
The uncertainty in $\alpha_b$ due to the
statistical uncertainties in $\langle |Q_{dif}|^2 \rangle$
and $\sigma_{sum}^2$ corresponds to a $1.6\%$ uncertainty in $A_b$.
The uncertainty in the hemisphere correlation parameter
$\lambda$ is estimated by varying fragmentation parameters
within JETSET 7.4, and by comparison with the 
HERWIG 5.7 \cite{HERWIG} fragmentation model.
The resulting uncertainty in $A_b$ is $ 1.4\%$. The
sensitivity of the result to the shape of the underlying
$Q_b$ distribution is tested by generating
various triangular distributions as well as double Gaussian
distributions with offset means. 
The test distributions are constrained to yield a $Q_{sum}$
distribution consistent with data,
and the total uncertainty
is found to be $0.8\%$. In addition,
while the mean value of the self-calibration
parameter $\alpha_b$ is constrained by the data, it 
has a $\cos\theta$ dependence due to the fall-off of the
tracking efficiency at high $|\cos\theta|$ which must be
estimated using the simulation, leading to a $0.4$\% uncertainty
in $A_b$.

The extracted value of $A_b$ is sensitive to our estimate
of the  
$Z^0 \rightarrow c {\overline c}$ background, which tends to reduce
the observed asymmetry due to the positive charge of the underlying
$c$ quark. 
The uncertainty in the purity estimate of $96.9 \pm 0.3\%$
is dominated by the uncertainties in the charm tagging efficiency
($\epsilon_c = 0.0218 \pm 0.0004$) and the statistical uncertainty
of the bottom tagging efficiency determined from data, leading to
a $0.5\%$ uncertainty in $A_b$. 
An outline of the charmed quark efficiency uncertainty determination
can be found in Ref.~\cite{cqeff}; the uncertainty is dominated by
empirical constraints on
charmed hadron production rates and on $K^0$ production
in the decay of charmed mesons. Uncertainties in the measured values
of $R_b$ and $R_c$ contribute, through the tag purity, to uncertainties
in $A_b$ of 0.1\% and 0.0\%, respectively.

Agreement between the data and MC simulation charged track multiplicity
distributions 
is obtained only after the inclusion of additional ad-hoc
tracking inefficiency. This random inefficiency was
parameterized as a function of total track momentum,
and averages 0.4 charged tracks per event, leading to
an overall change of $+1.3\%$ in $A_b$.
As a check, we employ
an alternative approach, matching the efficiency of the linking
of the independent CDC and VXD3 track segments between data and MC
simulation. This yields a change
of $+0.5\%$ in $A_b$; we take the difference of $0.8\%$ as an estimate
of the systematic error on the modeling of the tracking efficiency.
Combining all systematic uncertainties in quadrature yields
a total relative systematic uncertainty of $2.6\%$.

The extracted value of $A_b$ depends on a number of model parameters,
as follows. Increases by 0.01 in the values of $A_c$, $R_b$, $R_c$,
and the per-event rate of $b {\bar b}$ production
via gluon splitting, lead to changes in $A_b$ of
+0.0002, -0.0055, +0.0002, and +0.0110, respectively.

In conclusion, we have exploited the highly polarized SLC electron
beam and precise vertexing capabilities of the SLD detector
to perform a direct measurement of
 $A_b=0.906\pm 0.022(\mbox{stat})\pm 0.023 (\mbox{syst})$,
from the 1996-98 SLD data sample. Combined with our
previously published result~\cite{prl98} based on the 1993-95 data sample, we find
\begin{equation}
  A_b=0.907\pm 0.020(\mbox{stat})\pm 0.024 (\mbox{syst}),
\end{equation}
for the full 1993-98 data sample. This result is
in good agreement with the Standard Model prediction of 0.935,
and represents an
improvement of over a factor of two in the precision of the determination
of $A_b$ via the use of momentum-weighted track charge.

We thank the staff of the SLAC accelerator department for their
outstanding efforts on our behalf. This work was supported by the
U.S. Department of Energy and National Science Foundation, the UK Particle
Physics and Astronomy Research Council, the Istituto Nazionale di Fisica
Nucleare of Italy and the Japan-US Cooperative Research Project on High
Energy Physics.

%

%
%

\vfill
\eject

%
%

\begin{table}
\begin{center}
\begin{tabular}{lcc}
{\bf Error Source} & {\bf Variation} & $\delta A_b/A_b$ \\ \hline
 & & \\
\multicolumn{3}{l}{\it \underline{Self-Calibration}} \\ 
$\alpha_b$ statistics & $\pm$1$\sigma$ & 1.6\% \\
$\lambda_b$ Correlation & JETSET, HERWIG & 1.4\% \\
$P(Q_b)$ shape & Different shapes & 0.8\% \\
$\cos\theta$ shape of $\alpha_b$ & MC Shape $vs$ Flat & 0.4\% \\
Light Flavor  & 50\% of correction & 0.2\% \\
 & & \\
\multicolumn{3}{l}{\it \underline{Analysis}} \\
Tag Composition & Procedure from \cite{cqeff} & 0.5\% \\
Detector Modeling & Compare tracking & 0.8\% \\
 &eff. corrections& \\
Beam Polarization & $\pm$0.5\% & 0.5\% \\
QCD & Full correction& 0.3\% \\
Gluon Splitting &Full Correction& 0.1\% \\
$A_c$ & $0.67\pm 0.04$ & 0.1\% \\
$A_{bckg}$ & $0\pm 0.50$ & 0.2\% \\
 \hline
{\bf Total} & & {\bf 2.6\%}
\end{tabular}
\end{center}
\caption{Relative systematic errors on the measurement of $A_b$.}
\label{SYSERR}
\end{table}

\listoffigures
\listoftables

\vfill \eject

\begin{large}
\begin{center}
$^{**}$ {\bf The SLD Collaboration}
\end{center}
\end{large}

%
%
%
\begin{center}
\def\iAOMORI{$^{(1)}$}
\def\iBRI{$^{(2)}$}
\def\iBRUN{$^{(3)}$}
\def\iBU{$^{(4)}$}
\def\iCOLO{$^{(5)}$}
\def\iCSU{$^{(6)}$}
\def\iFERR{$^{(7)}$}
\def\iFRAS{$^{(8)}$}
\def\iJHU{$^{(9)}$}
\def\iLBL{$^{(10)}$}
\def\iMASS{$^{(11)}$}
\def\iMISSI{$^{(12)}$}
\def\iMIT{$^{(13)}$}
\def\iMOSCOW{$^{(14)}$}
\def\iNAGO{$^{(15)}$}
\def\iOREG{$^{(16)}$}
\def\iOXF{$^{(17)}$}
\def\iPERU{$^{(18)}$}
\def\iRAL{$^{(19)}$}
\def\iRUTG{$^{(20)}$}
\def\iSLAC{$^{(21)}$}
\def\iSOONG{$^{(22)}$}
\def\iTENN{$^{(23)}$}
\def\iTOHO{$^{(24)}$}
\def\iUCSB{$^{(25)}$}
\def\iUCSC{$^{(26)}$}
\def\iVAND{$^{(27)}$}
\def\iWASH{$^{(28)}$}
\def\iWISC{$^{(29)}$}
\def\iYALE{$^{(30)}$}

  \baselineskip=.75\baselineskip
\mbox{Kenji Abe\unskip,\iNAGO}
\mbox{Koya Abe\unskip,\iTOHO}
\mbox{T. Abe\unskip,\iSLAC}
\mbox{I. Adam\unskip,\iSLAC}
\mbox{H. Akimoto\unskip,\iSLAC}
\mbox{D. Aston\unskip,\iSLAC}
\mbox{K.G. Baird\unskip,\iMASS}
\mbox{C. Baltay\unskip,\iYALE}
\mbox{H.R. Band\unskip,\iWISC}
\mbox{T.L. Barklow\unskip,\iSLAC}
\mbox{J.M. Bauer\unskip,\iMISSI}
\mbox{G. Bellodi\unskip,\iOXF}
\mbox{R. Berger\unskip,\iSLAC}
\mbox{G. Blaylock\unskip,\iMASS}
\mbox{J.R. Bogart\unskip,\iSLAC}
\mbox{G.R. Bower\unskip,\iSLAC}
\mbox{J.E. Brau\unskip,\iOREG}
\mbox{M. Breidenbach\unskip,\iSLAC}
\mbox{W.M. Bugg\unskip,\iTENN}
\mbox{D. Burke\unskip,\iSLAC}
\mbox{T.H. Burnett\unskip,\iWASH}
\mbox{P.N. Burrows\unskip,\iOXF}
\mbox{A. Calcaterra\unskip,\iFRAS}
\mbox{R. Cassell\unskip,\iSLAC}
\mbox{A. Chou\unskip,\iSLAC}
\mbox{H.O. Cohn\unskip,\iTENN}
\mbox{J.A. Coller\unskip,\iBU}
\mbox{M.R. Convery\unskip,\iSLAC}
\mbox{V. Cook\unskip,\iWASH}
\mbox{R.F. Cowan\unskip,\iMIT}
\mbox{G. Crawford\unskip,\iSLAC}
\mbox{C.J.S. Damerell\unskip,\iRAL}
\mbox{M. Daoudi\unskip,\iSLAC}
\mbox{N. de Groot\unskip,\iBRI}
\mbox{R. de Sangro\unskip,\iFRAS}
\mbox{D.N. Dong\unskip,\iSLAC}
\mbox{M. Doser\unskip,\iSLAC}
\mbox{R. Dubois\unskip,\iSLAC}
\mbox{I. Erofeeva\unskip,\iMOSCOW}
\mbox{V. Eschenburg\unskip,\iMISSI}
\mbox{S. Fahey\unskip,\iCOLO}
\mbox{D. Falciai\unskip,\iFRAS}
\mbox{J.P. Fernandez\unskip,\iUCSC}
\mbox{K. Flood\unskip,\iMASS}
\mbox{R. Frey\unskip,\iOREG}
\mbox{E.L. Hart\unskip,\iTENN}
\mbox{K. Hasuko\unskip,\iTOHO}
\mbox{S.S. Hertzbach\unskip,\iMASS}
\mbox{M.E. Huffer\unskip,\iSLAC}
\mbox{X. Huynh\unskip,\iSLAC}
\mbox{M. Iwasaki\unskip,\iOREG}
\mbox{D.J. Jackson\unskip,\iRAL}
\mbox{P. Jacques\unskip,\iRUTG}
\mbox{J.A. Jaros\unskip,\iSLAC}
\mbox{Z.Y. Jiang\unskip,\iSLAC}
\mbox{A.S. Johnson\unskip,\iSLAC}
\mbox{J.R. Johnson\unskip,\iWISC}
\mbox{R. Kajikawa\unskip,\iNAGO}
\mbox{M. Kalelkar\unskip,\iRUTG}
\mbox{H.J. Kang\unskip,\iRUTG}
\mbox{R.R. Kofler\unskip,\iMASS}
\mbox{R.S. Kroeger\unskip,\iMISSI}
\mbox{M. Langston\unskip,\iOREG}
\mbox{D.W.G. Leith\unskip,\iSLAC}
\mbox{V. Lia\unskip,\iMIT}
\mbox{C. Lin\unskip,\iMASS}
\mbox{G. Mancinelli\unskip,\iRUTG}
\mbox{S. Manly\unskip,\iYALE}
\mbox{G. Mantovani\unskip,\iPERU}
\mbox{T.W. Markiewicz\unskip,\iSLAC}
\mbox{T. Maruyama\unskip,\iSLAC}
\mbox{A.K. McKemey\unskip,\iBRUN}
\mbox{R. Messner\unskip,\iSLAC}
\mbox{K.C. Moffeit\unskip,\iSLAC}
\mbox{T.B. Moore\unskip,\iYALE}
\mbox{M. Morii\unskip,\iSLAC}
\mbox{D. Muller\unskip,\iSLAC}
\mbox{V. Murzin\unskip,\iMOSCOW}
\mbox{S. Narita\unskip,\iTOHO}
\mbox{U. Nauenberg\unskip,\iCOLO}
\mbox{H. Neal\unskip,\iYALE}
\mbox{G. Nesom\unskip,\iOXF}
\mbox{N. Oishi\unskip,\iNAGO}
\mbox{D. Onoprienko\unskip,\iTENN}
\mbox{L.S. Osborne\unskip,\iMIT}
\mbox{R.S. Panvini\unskip,\iVAND}
\mbox{C.H. Park\unskip,\iSOONG}
\mbox{I. Peruzzi\unskip,\iFRAS}
\mbox{M. Piccolo\unskip,\iFRAS}
\mbox{L. Piemontese\unskip,\iFERR}
\mbox{R.J. Plano\unskip,\iRUTG}
\mbox{R. Prepost\unskip,\iWISC}
\mbox{C.Y. Prescott\unskip,\iSLAC}
\mbox{B.N. Ratcliff\unskip,\iSLAC}
\mbox{J. Reidy\unskip,\iMISSI}
\mbox{P.L. Reinertsen\unskip,\iUCSC}
\mbox{L.S. Rochester\unskip,\iSLAC}
\mbox{P.C. Rowson\unskip,\iSLAC}
\mbox{J.J. Russell\unskip,\iSLAC}
\mbox{O.H. Saxton\unskip,\iSLAC}
\mbox{T. Schalk\unskip,\iUCSC}
\mbox{B.A. Schumm\unskip,\iUCSC}
\mbox{J. Schwiening\unskip,\iSLAC}
\mbox{V.V. Serbo\unskip,\iSLAC}
\mbox{G. Shapiro\unskip,\iLBL}
\mbox{N.B. Sinev\unskip,\iOREG}
\mbox{J.A. Snyder\unskip,\iYALE}
\mbox{H. Staengle\unskip,\iCSU}
\mbox{A. Stahl\unskip,\iSLAC}
\mbox{P. Stamer\unskip,\iRUTG}
\mbox{H. Steiner\unskip,\iLBL}
\mbox{D. Su\unskip,\iSLAC}
\mbox{F. Suekane\unskip,\iTOHO}
\mbox{A. Sugiyama\unskip,\iNAGO}
\mbox{S. Suzuki\unskip,\iNAGO}
\mbox{M. Swartz\unskip,\iJHU}
\mbox{F.E. Taylor\unskip,\iMIT}
\mbox{J. Thom\unskip,\iSLAC}
\mbox{E. Torrence\unskip,\iMIT}
\mbox{T. Usher\unskip,\iSLAC}
\mbox{J. Va'vra\unskip,\iSLAC}
\mbox{R. Verdier\unskip,\iMIT}
\mbox{D.L. Wagner\unskip,\iCOLO}
\mbox{A.P. Waite\unskip,\iSLAC}
\mbox{S. Walston\unskip,\iOREG}
\mbox{A.W. Weidemann\unskip,\iTENN}
\mbox{E.R. Weiss\unskip,\iWASH}
\mbox{J.S. Whitaker\unskip,\iBU}
\mbox{S.H. Williams\unskip,\iSLAC}
\mbox{S. Willocq\unskip,\iMASS}
\mbox{R.J. Wilson\unskip,\iCSU}
\mbox{W.J. Wisniewski\unskip,\iSLAC}
\mbox{J.L. Wittlin\unskip,\iMASS}
\mbox{M. Woods\unskip,\iSLAC}
\mbox{T.R. Wright\unskip,\iWISC}
\mbox{R.K. Yamamoto\unskip,\iMIT}
\mbox{J. Yashima\unskip,\iTOHO}
\mbox{S.J. Yellin\unskip,\iUCSB}
\mbox{C.C. Young\unskip,\iSLAC}
\mbox{H. Yuta\unskip.\iAOMORI}

\it
  \vskip \baselineskip                   
  \centerline{(The SLD Collaboration)}   
  \vskip \baselineskip
  \baselineskip=.75\baselineskip   
\iAOMORI
  Aomori University, Aomori , 030 Japan, \break
\iBRI
  University of Bristol, Bristol, United Kingdom, \break
\iBRUN
  Brunel University, Uxbridge, Middlesex, UB8 3PH United Kingdom, \break
\iBU
  Boston University, Boston, Massachusetts 02215, \break
\iCOLO
  University of Colorado, Boulder, Colorado 80309, \break
\iCSU
  Colorado State University, Ft. Collins, Colorado 80523, \break
\iFERR
  INFN Sezione di Ferrara and Universita di Ferrara, I-44100 Ferrara, Italy, \break
\iFRAS
  INFN Lab. Nazionali di Frascati, I-00044 Frascati, Italy, \break
\iJHU
  Johns Hopkins University,  Baltimore, Maryland 21218-2686, \break
\iLBL
  Lawrence Berkeley Laboratory, University of California, Berkeley, California 94720, \break
\iMASS
  University of Massachusetts, Amherst, Massachusetts 01003, \break
\iMISSI
  University of Mississippi, University, Mississippi 38677, \break
\iMIT
  Massachusetts Institute of Technology, Cambridge, Massachusetts 02139, \break
\iMOSCOW
  Institute of Nuclear Physics, Moscow State University, 119899, Moscow Russia, \break
\iNAGO
  Nagoya University, Chikusa-ku, Nagoya, 464 Japan, \break
\iOREG
  University of Oregon, Eugene, Oregon 97403, \break
\iOXF
  Oxford University, Oxford, OX1 3RH, United Kingdom, \break
\iPERU
  INFN Sezione di Perugia and Universita di Perugia, I-06100 Perugia, Italy, \break
\iRAL
  Rutherford Appleton Laboratory, Chilton, Didcot, Oxon OX11 0QX United Kingdom, \break
\iRUTG
  Rutgers University, Piscataway, New Jersey 08855, \break
\iSLAC
  Stanford Linear Accelerator Center, Stanford University, Stanford, California 94309, \break
\iSOONG
  Soongsil University, Seoul, Korea 156-743, \break
\iTENN
  University of Tennessee, Knoxville, Tennessee 37996, \break
\iTOHO
  Tohoku University, Sendai 980, Japan, \break
\iUCSB
  University of California at Santa Barbara, Santa Barbara, California 93106, \break
\iUCSC
  University of California at Santa Cruz, Santa Cruz, California 95064, \break
\iVAND
  Vanderbilt University, Nashville,Tennessee 37235, \break
\iWASH
  University of Washington, Seattle, Washington 98105, \break
\iWISC
  University of Wisconsin, Madison,Wisconsin 53706, \break
\iYALE
  Yale University, New Haven, Connecticut 06511. \break

\rm
%

\end{center}

\end{document}